\documentclass[sigconf]{acmart}
\usepackage{balance}
\usepackage{enumitem}
\usepackage{caption}
\usepackage{multirow}
\usepackage{hyperref}
\usepackage[ruled,vlined,linesnumbered]{algorithm2e}
\usepackage{times}

\usepackage{graphicx} 
\usepackage{subfigure}
\usepackage{longtable}
\usepackage{arydshln}
\usepackage{booktabs}
 \usepackage{microtype}
 
\usepackage{xcolor}
\usepackage{colortbl}
\usepackage{makecell}

\AtBeginDocument{%
  \providecommand\BibTeX{{%
    \normalfont B\kern-0.5em{\scshape i\kern-0.25em b}\kern-0.8em\TeX}}}

\copyrightyear{2023}
\acmYear{2023}
\setcopyright{acmlicensed}
\acmConference[MM '23]{Proceedings of the 31st ACM International Conference on Multimedia}{October 29-November 3, 2023}{Ottawa, ON, Canada}
\acmBooktitle{Proceedings of the 31st ACM International Conference on Multimedia (MM '23), October 29-November 3, 2023, Ottawa, ON, Canada}
\acmPrice{15.00}
\acmDOI{10.1145/3581783.3611826}
\acmISBN{979-8-4007-0108-5/23/10}

\newcommand{\modelname}{ COPA }
\newcommand{\PretrainTaskName}{ Patch-Text Alignment }

\begin{document}
	
	\title{\modelname: Efficient Vision-Language Pre-training Through Collaborative Object- and Patch-Text Alignment}
	
	

	\author{Chaoya Jiang}
	\email{jiangchaoya@pku.edu.cn}
	\affiliation{%
		\institution{National Engineering Research Center for Software Engineering, Peking University, Peking University}
		 \city{Beijing}
		 \country{China}
	}

	\author{Haiyang Xu}
	\authornote{Corresponding authors.}
	\email{shuofeng.xhy@alibaba-inc.com}
	\affiliation{%
		\institution{DAMO Academy, Alibaba Group}
		\city{Hangzhou}
		 \country{China}
	}
	
	\author{Wei Ye}
        \authornotemark[1]
	\email{wye@pku.edu.cn}
	\affiliation{%
		\institution{National Engineering Research Center for Software Engineering, Peking University, Peking University}
		 \city{Beijing}
		 \country{China}
	}

	\author{Qinghao Ye}
	\email{yeqinghao.yqh@alibaba-inc.com}
	\affiliation{%
		\institution{DAMO Academy, Alibaba Group}
		\city{Hangzhou}
		\country{China}
	}
        
        \author{Chenliang Li}
	\email{lcl193798@alibaba-inc.com}
	\affiliation{%
		\institution{DAMO Academy, Alibaba Group}
		\city{Hangzhou}
		\country{China}
	}
	
	\author{Ming Yan}
	\email{ym119608@alibaba-inc.com}
	
	\affiliation{%
		\institution{DAMO Academy, Alibaba Group}
		\city{Hangzhou}
		\country{China}
	}

        \author{Bin Bi}
	\email{b.bi@alibaba-inc.com}
	
	\affiliation{%
		\institution{DAMO Academy, Alibaba Group}
		\city{Hangzhou}
		\country{China}
	}
        \author{Shikun Zhang}
         
	\email{zhangsk@pku.edu.cn}
	\affiliation{%
		\institution{National Engineering Research Center for Software Engineering, Peking University, Peking University}
		 \city{Beijing}
		 \country{China}
	}
	
	\author{Ji Zhang}
	\email{zj122146@alibaba-inc.com}
	\affiliation{%
		\institution{DAMO Academy, Alibaba Group}
		\city{Hangzhou}
		\country{China}
	}
 
	\author{Fei Huang}
	\email{f.huang@alibaba-inc.com}
	
	\affiliation{%
		\institution{DAMO Academy, Alibaba Group}
		\city{Hangzhou}
		\country{China}
	}


\renewcommand{\shortauthors}{Chaoya Jiang et al.}

	\begin{abstract}
		    Vision-Language Pre-training (VLP) methods based on object detection enjoy the rich knowledge of fine-grained object-text alignment but at the cost of computationally expensive inference. Recent Visual-Transformer (ViT)-based approaches circumvent this issue while struggling with long visual sequences without detailed cross-modal alignment information. This paper introduces a ViT-based VLP technique that efficiently incorporates object information through a novel patch-text alignment mechanism. Specifically, we convert object-level signals into patch-level ones and devise a \PretrainTaskName pre-training task (PTA) to learn a text-aware patch detector. By using off-the-shelf delicate object annotations in 5\% training images, we jointly train PTA with other conventional VLP objectives in an end-to-end manner, bypassing the high computational cost of object detection and yielding an effective patch detector that accurately detects text-relevant patches, thus considerably reducing patch sequences and accelerating computation within the ViT backbone. Our experiments on a variety of widely-used benchmarks reveal that our method achieves a speedup of nearly 88\% compared to prior VLP models while maintaining competitive or superior performance on downstream tasks with similar model size and data scale.
	\end{abstract}

\ccsdesc[500]{Information systems~Multimedia and multimodal retrieval}
\ccsdesc[500]{Computing methodologies~Image representations}
\ccsdesc[300]{Computing methodologies~Object detection}
	\keywords{Vision-Language Pretraining; Efficiency; Patch-Text Alignment; Detection}


	\maketitle

	\section{Introduction}

 \begin{figure}[!t]
\centering
\includegraphics[width=0.48\textwidth]{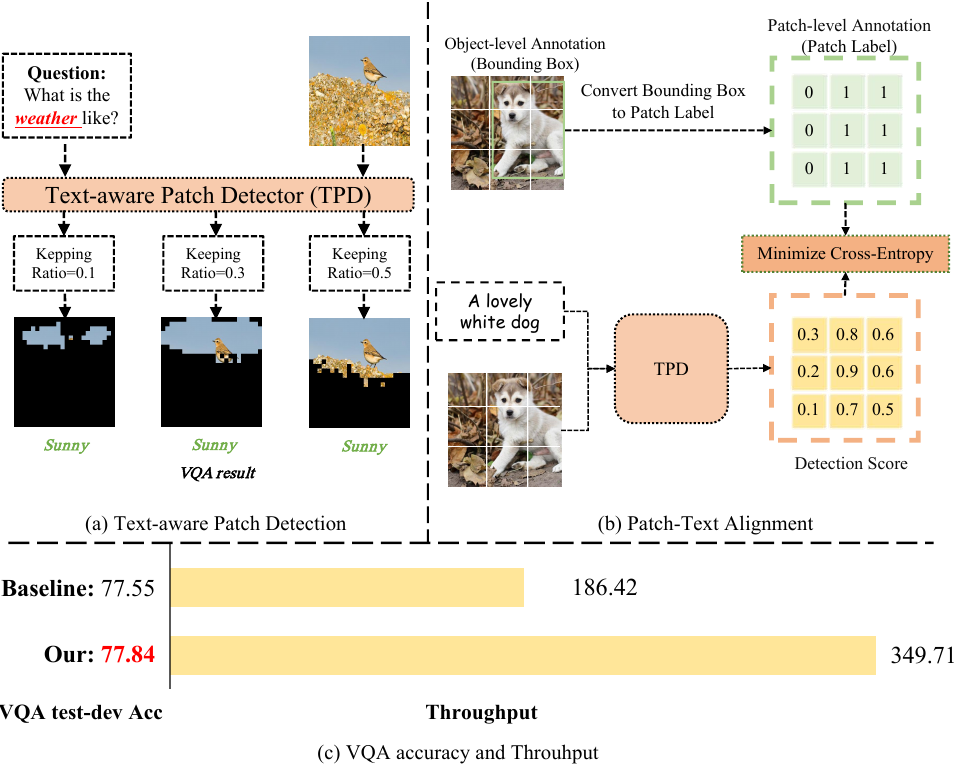}

\caption{Subfigure (a) illustrates the impact of Text-aware Patch Detector (TPD) in the VQA scenario on various keeping ratios, which is a hyperparameter determining the proportion of retained visual tokens to all tokens. Subfigure (b) demonstrates how Patch-Text Alignment converts object-level annotations to patch-level annotations and optimizes TPD based on the obtained supervision signals. Subfigure (c) presents the VQA accuracy and throughput results for our VLP model and the baseline.}
\label{fig:fig1}
\vspace{-4ex}
\end{figure}

Recently, Vision-Language Pre-training (VLP)~\cite{Tan2019LXMERTLC, Chen2020UNITERUI,Lu2019ViLBERTPT,Huang2020PixelBERTAI,Su2020VLBERTPO, Li2020OscarOA,Chen2020UNITERUI,Zhou2020UnifiedVP,Li2021AlignBF,Yu2021ERNIEViLKE,li2022mplug} has achieved remarkable success across a wide range of Vision-Language (VL) tasks, establishing itself as a dominant paradigm. Existing VLP methods can be broadly classified into two categories based on their approach to image feature extraction.

\textbf{Detection-based Models}~\cite{Tan2019LXMERTLC, Chen2020UNITERUI,Lu2019ViLBERTPT,Li2020OscarOA} identify objects/regions within images by employing pre-trained object detectors \cite{Ren2015FasterRT,Redmon2016YouOL,He2017MaskR} and learn to align detected objects/regions with text. Although these methods benefit from the valuable fine-grained cross-modal alignment knowledge, they are burdened by the considerable computational cost of object/region detection and the error propagation issues arising from the two-step pre-training strategy.

\textbf{ViT-based Models} refer to more recent efforts~\cite{Li2021AlignBF,Radford2021LearningTV,Kim2021ViLTVT,Wang2021VLMoUV,Singh2021FLAVAAF,li2022mplug,Jiang2023BUSE,Jiang2022TRIPSEV,Jiang2023VisionLP,Jiang2023TiMixTI,Jiang2023ExploitingPI} built on Vision Transformer (ViT)~\cite{Dosovitskiy2021AnII}. They utilize ViT as the visual encoder or cross-modal fusion encoder due to its capacity to handle lengthy visual sequences derived from image grids/patches. Although these approaches sidestep the high computational cost and error propagation associated with object detection, they still face two challenges. First, they can only learn coarse-grained alignments between the entire image and text, owing to the absence of fine-grained alignment annotations (e.g., between patches and text), while acquiring fine-grained alignment in VLP is essential for numerous cross-modal understanding and reasoning tasks (e.g., Visual Question Answering and Visual Grounding). Second, many redundant patches in the image do not align with the input text, resulting in extended visual sequences, particularly for high-resolution images. For instance, as illustrated in Figure \ref{fig:fig1}(a), during the inference of the VQA task with an image size of 386$\times$386 for the question "What is the weather like?", approximately 70\% of the patches in the image depict a bird and sand, which do not correspond to the question. Eliminating these text-irrelevant patches will not impact the model output but can significantly reduce inference time.

To harness the strengths of both model types while mitigating their shortcomings, we propose a novel VLP method named \textbf{COPA}, which effectively incorporates fine-grained object-text alignment knowledge into a ViT-based model in a lightweight and efficient manner. COPA stands for \textbf{C}ollaboration between \textbf{O}bject- and \textbf{P}atch-Text \textbf{A}lignment. Specifically, by transforming object/region annotations to patch-level annotations, we design an innovative auxiliary pre-training task—\PretrainTaskName (PTA)—to capture fine-grained cross-modal alignment knowledge and train an accurate Text-aware Patch Detector (TPD). Our method offers two key benefits. First, PTA is jointly trained with other traditional VLP pre-training tasks using ready-made object annotations from just 5\% of training images, resulting in an end-to-end architecture without the need for additional computation, e.g., of object detection in prior detection-based models. Second, guided by PTA, the TPD, which is integrated into the ViT-based visual backbone, can accurately identify text-related patches, thereby reducing the length of the visual sequence and further decreasing the overall computational cost. Impressively, we employ a relatively small number of object-level labels from the COCO~\cite{Lin2014MicrosoftCC}(0.11M) and VG~\cite{Krishna2016VisualGC}(0.10M) datasets, yet succeed in training a robust and generalizable patch detector that can effectively identify text-relevant patches in large-scale and possibly out-of-domain pre-training data in CC~\cite{cc}(3M) dataset (see Figure~\ref{fig:acc_recall_TPD}).

We evaluate \modelname on various representative VL understanding and generation tasks, including visual question answering, cross-modal retrieval, and image captioning. Our findings show that by retaining only 50\% text-relevant image patches in the visual backbone, we can achieve competitive or even superior downstream task performance. Simultaneously, the efficiency (more specifically, the throughput) of \modelname increases by 88\% compared to previous similar VLP models. For example, as illustrated in Figure~\ref{fig:fig1} (c) and Table~\ref{table_efficient}, \modelname boosts the throughput of the baseline from 186.42 to 349.71 and even improves by about 0.3 on the VQA test-dev under the same experimental settings. Moreover, by increasing the input image resolution, \modelname attains well-designed state-of-the-art downstream task performance (e.g., 78.25 on VQA test-dev, see Table~\ref{table: image_size}) as it benefits from incorporating more image tokens without raising computational costs compared to other baselines.

\begin{figure*}[h]
\centering


\centering
\includegraphics[width=0.95\linewidth]{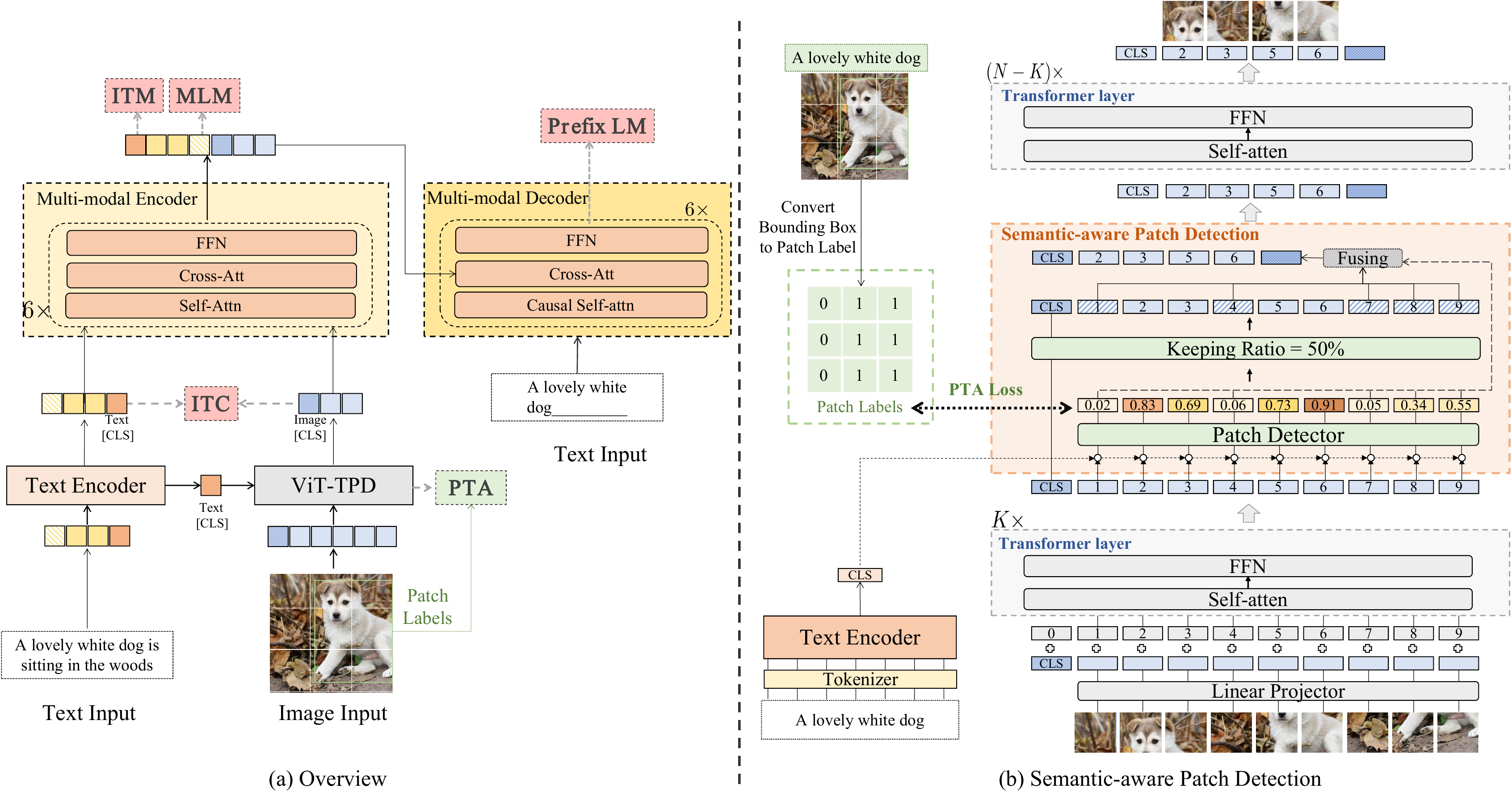}

\caption{(a) Overview of our VLP model (\modelname). By incorporating the PTA task, we can learn the fine-grained patch-text alignment end-to-end through joint optimization with other pre-training tasks. (b) The architecture of the Text-aware Patch Detector (TPD) is plugged into the ViT-base visual backbone (ViT-TPD). In this sub-figure, we give a simplified example to show how to detect text-relevant patches and calculate the PTA Loss.}
\label{pipline}
\vspace{-2ex}
\end{figure*}

\section{Related Work}
\subsection{Vision-Language Pre-training} 
The existing work on vision language pre-training typically falls into two categories: Detector-based VLP model and CNN/ViT-Based VLP models.
Previous Detector-based VLP methods \cite{Lu2019ViLBERTPT,Li2019VisualBERTAS, Tan2019LXMERTLC, Li2020OscarOA,Chen2020UNITERUI,Yu2021ERNIEViLKE} mainly take a two-step training pipeline approach, which first extracts visual features by a pre-trained object detector and then trains the cross-modal pre-training model to align text and visual features. Even though there are some region-based methods that reduce the computation cost with the lightweight model architecture \cite{Wang2020MiniVLMAS}, those methods still suffer from the following weaknesses: (1) the expensive computational cost and time consumption for detecting objects/regions. (2) the error propagation problems caused by the two-step pre-training strategy. More recently, ViTs-based \cite{Li2021AlignBF,Kim2021ViLTVT,Radford2021LearningTV,Wang2021VLMoUV,li2022blip,li2022mplug, Wang2021SimVLMSV,Kim2021ViLTVT} methods (especially the patch-based ViT)  removes the complicated object detector in feature extraction to conduct end-to-end VL learning. These methods avoid the drawbacks of object detectors but face excessively long visual sequences without fine-grained cross-modal alignment information. Such long visual sequences also bring expensive computation costs but there is no work that focuses on decreasing the high computational cost. In this work, we propose a novel method to detect the text-relevant patches and reduce the redundant undetected patches to a single one in the backbone, thus can decrease the computation cost of VLP models.

\subsection{ViTs Acceleration}

To accelerate the computation of the transformer\cite{Vaswani2017AttentionIA} based model, many studies focus on proposing more efficient attention mechanisms \cite{Wang2020LinformerSW, Kitaev2020ReformerTE, Choromanski2021RethinkingAW} or compress Transformer structures \cite{Liu2021SwinTH, Heo2021RethinkingSD,Wang2021PyramidVT}. Recently, some approaches have focused on accelerating ViTs by reducing the number of tokens involved in the inference of ViTs. For example, to expedite ViTs, \citet{Ryoo2021TokenLearnerAS} proposed TokenLearner, in which a relatively small amount of tokens are learned by aggregating the entire feature map weighted by dynamic attention. \citet{Rao2021DynamicViTEV} introduces a method to reduce tokens for a fully trained ViT, where an extra learnable neural network is added to ViT to select a subset of tokens. \citet{Liang2022NotAP} proposes to reduce the computational overhead of inference by proposing a token reorganization method to reduce and reorganize image tokens progressively. However, those methods are unsuitable for VLP as they reduce the image tokens without considering the text context. There is recent VLP work~\cite{Jiang2022TRIPSEV} has noticed this issue and attempted to reduce the redundant visual tokens based on coarse-grained cross-modal semantic alignment. However, due to the absence of fine-grained visual-text supervision, they struggle to accurately select text-relevant patches, thereby leading to a degradation in the model performance.  
 
\section{Method}
In this section, we will first give an overview of our model architecture and then introduce the Text-aware Patch Detector (TPD) in the Vision Transformer (ViT) backbone. Finally, we give the details about the pre-training task of \PretrainTaskName (PTA).

\subsection{Model Architecture}
As shown in Figure~\ref{pipline}(a), \modelname contains a visual encoder, a text encoder, a multimodal fusion encoder for performing cross-modal interaction, and a multimodal decoder for text generation (Note that we implement our method based on \citet{li2022mplug} which provides details about the model architecture). The visual encoder named ViT-TPD is a ViT-based network that consists of multiple Transformer layers and a Text-aware Patch Detector (TPD) utilized to detect text-relevant patches.

Formally, suppose we have an input image-text pair denoted as $\left(I, T\right)$. For the input text, we feed it to the text encoder and get the text representation $T = \{t_{cls}, t_1, t_2, \cdots, t_m \}$ where $t_{cls}$ is the embedding of the text [CLS] token which is used to summarize the global semantic information of the text. For the input image, we divide the input image into $n$ non-overlapping patches $P = \{p_{cls}, p_1,p_2, \cdots, p_{n} \}$. Then we feed the patch sequence to the visual encoder ViT-TPD and get the patch sequence representation $V=\{v_{cls}, v_1, v_2, \cdots,v_u\}, u\textless n $. In the ViT-TPD, we apply the text-aware patch detector to detect the text-relevant image patches and fuse other undetected redundant patches to a single token, which can reduce the visual sequence length for training and inference efficiency. After that, the image and text representations are fed into the cross-modal encoder and we get the cross-modal representations $\{c_{cls}, c_1, c_2, \dots, c_{l}\}$ where $l=u+m$. The cross-modal representations can be used to finetune downstream multi-modal understanding tasks. Besides, the output cross-modal representations $\{c_{cls}, c_1, c_2, \dots, c_{l}\}$ of the multi-modal encoder are fed into a Transformer decoder for sequence-to-sequence learning. 

 
\subsection{Text-aware Patch Detection}

As shown in Figure \ref{pipline}, our ViT-based visual backbone contains $N$ standard Transformer layers and a plug-and-play Text-aware Patch Detector (TPD), which is the only difference from the standard ViT. The TPD dynamically detects the image patches with the guidance of textual input. Specially, suppose the TPD is plugged between the $k_{th}$ ( $1\leq k  \textless N$ ) Transformer layer and $(k+1)_{th}$ Transformer layer. Suppose the output patch sequence features of $k_{th}$ Transformer layer is $v^{k}=\{v^{k}_{cls}, v^{k}_1, \cdots, v^{k}_n\}$. We exclude the image [CLS] tokens $v^k_{cls}$ and feed the left patch tokens $\{v^{k}_1, \cdots, v^{k}_n\}$ and the text [CLS] feature $t_{cls}$ together to the TPD. The text [CLS] feature is output by the text encoder and represents global information of the input text $T$.

In the TPD, we first concatenate the text [CLS] feature with each image patch token as follows:
$$
    \dot{v}^{k}_i = concat(v^{k}_i, t_{cls})
$$
where $v^{k}_i\in R^d, t_{cls}\in R^d, \dot{v}^{k}_i\in R^{2d}, i \in \{1, 2, \dots, n \}$. Then the concatenated patch features $\{\dot{v}^{k}_i\}$ are fed to the patch detector. The patch detector is an MLP that contains three linear layers and is used to predict the alignment score between patches and the input text T. The first two linear layers will linearly project the concatenated patch features $\{\dot{v}^{k}_i\}$ to the hidden representations $\{h^{k}_i\}$ and then the hidden representations $\{h^{k}_i\}$ is fed to the last linear layer denoted as $\mathbf{F}_{\theta}$ which can be seen as a classifier to predict whether the patches are relevant to the input text. The output of the last linear layer has only one dimension and will be fed to a Sigmoid activation function. Formally, the alignment score $a_i$ between the $i_{th}$ image patch and input text T can be calculated as follow:
$$
    a_{i} = Sigmoid(\mathbf{F}_{\theta}(h^{k}_i )), i \in \{1,2,\dots, n\}
$$
 Then, we identify and preserve the image tokens with high alignment scores with the text which corresponds to the $K$ largest elements in the alignment score sequence $\{a_{1}, ..a_{n}\}$, where $K = n \times \alpha$, and $ \alpha$ is a hyper-parameter and named Keeping Ratio which is used to control the proportion of detected patches to total patches.  
The detected top-K image patch tokens are kept and the undetected patch tokens $\{v_{z_1}, v_{z_2}, \cdots,v_{z_{n-K}}\}$ which generally have lower alignment scores with the text will be treated as text-irrelevant tokens and further fused by a token fusion operation. We fuse undetected tokens to one token $v_{f}$ by a weighted sum operation to supplement one as follows:
\begin{equation}
    \left[\hat{a}_{z_1}, \cdot, \hat{a}_{z_{n-k}} \right] = Softmax(\left[{a}_{z_1}, \cdot, {a}_{z_{n-k}} \right] )
\end{equation}
\begin{equation}
    v_{f} = \sum\limits^{n-k}\limits_{i = 1} \hat{a}_{z_i} \cdot v_{z_i}
\end{equation}
    
After fusing the undetected patch tokens to single one token $v_{f}$,  we reconstruct the $k_{th}$ visual sequence as $v^{k} = \left[v^{k}_{cls},v^{k}_{1}, \cdots,v^{k}_{u},{v}^{k}_{f}\right]$, which consists of the image [CLS] token embedding, the detected text-relevant image patch embeddings, and the fused patch embedding. Then the reduced visual sequence is fed to the next $(k+1)_{th}$ transformer layer.
Such Text-aware Patch Detector (TPD) works during both pre-training and finetuning, which can be optimized in the pre-training by the \PretrainTaskName objectives which will be introduced in the next.
 
\subsection{\PretrainTaskName}
\label{subsection:PTA}
The key component of \modelname is the text-aware Patch Detector which needs to detect text-relevant patches according to the fine-grained alignment scores between the image patches and input text. However, such fine-grained patch-text alignment capabilities of traditional ViT-based models are weak as the lack of fine-grained patch-text labels. To address the above difficulties, in this sub-section, we introduce a novel pre-training task named \PretrainTaskName which facilitates the patch detector training and drives our model to learn the fine-grained patch-text alignment.

We find that in most object objection and visual grounding datasets, the object and region generally be paired with a class label or text description. Therefore, we can transfer every object class label to a text description based on a text template such as "This is a [class label].". Thus, for each (object/region) bounding box in an image, we can find a text description. Then, we transform the bounding box annotations to the patch-level labels by following this rule: Given an image and a bounding box annotation, if there is an overlap between an image patch and a bounding box, it will be assigned with label 1, otherwise, it will be assigned with label 0. For different text descriptions and bounding boxes, the labels of the patch are different.
In this way, we can generate fine-grained patch-text labels which can be served as the supervisory signal to pre-train our model.

After that, in each step of pre-training, we randomly sample a mini-batch of images from the object detection/visual grounding datasets (e.g., COCO~\cite{Lin2014MicrosoftCC} or VG~\cite{Krishna2016VisualGC}).  For each image, we randomly select an object/region bounding box and translate the bounding box annotation to the image patch label sequence following the transformation rule we mentioned before. Then, we feed the batch of text descriptions of the bounding boxes and the images together to our VLP model. In the ViT-TPD backbone, we hope the text-aware Patch Detector can detect all patches which have overlap with the region in the bounding box with the guidance of the bounding box text description. 
Supposing the Text-aware Patch Detector has predicted the alignment scores between image patches and text, we will calculate the binary cross entropy loss between the alignment scores and patch labels as follows:
    \begin{equation}
        \mathbf{L}_{PTA} = \frac{1}{e}\sum_{i=1}^{e} Y_{i} log\left(a_i\right) + \left(1-Y_{i}\right)log\left(1-a_i\right) 
    \end{equation}
where $a_i$ is the alignment score between $i_{th}$ patch in the image and the input text, $Y_i$ is the patch label of $i_{th}$ patch.
After calculating the \PretrainTaskName loss $\mathbf{L}_{PTA}$, we then randomly sample a mini-batch of normal image-text pairs from the dataset of 4M images (refer to subsection~\ref{data_setup}) and calculate the Image-Text Contrastive (ITC) loss $\mathbf{L}_{ITC}$, Image-Text Matching (ITM) loss $\mathbf{L}_{ITM}$, Masked Language Modeling (MLM) loss $\mathbf{L}_{MLM}$ and Prefix Language Modeling (PrefixLM) loss $\mathbf{L}_{Prefix}$ based on other four pre-training objectives (for more details about other pre-training objectives, please refer to Appendix~\ref{pretraing objectives}). We assign equal loss weights to each pre-training loss, and thus the full pre-training loss is:
\begin{equation}
     \mathbf{L} = \mathbf{L}_{ITC}+\mathbf{L}_{ITM} +\mathbf{L}_{MLM} + \mathbf{L}_{Prefix} + \mathbf{L}_{PTA}
\end{equation} 
We also provide the pseudo algorithm~\ref{alg:pretrain} to further elaborate on our pre-training schedule. Besides, at the beginning of pre-training, as the PTA loss has not yet converged, thus the performance of the patch detector is not ideal, we detect the image patches directly based on the attention weights of the image [CLS] token to other patch tokens. As the PTA loss gradually converges, we will use the patch detector to detect the text-relevant patches in the TPD module.
\vspace{-2ex}

\begin{algorithm}[htbp]
 \small
  
\caption{Pre-training algorithm of \modelname.}
\label{alg:pretrain}
  \KwIn{Large scale pre-training dataset $\mathcal{D}$, Object/Region Dataset $\mathcal{O}$, the number of pre-training epochs $T$, the pre-training learning rate $\alpha$, the batch size $B_D$ of dataset $\mathcal{D}$, the batch size $B_O$ of dataset $\mathcal{O}$.  }

  Initialize the parameters $\theta$ of our model $M$ \;
  \For{$t=1$ to $T$}{
    Randomly sample a mini-batch of $B_O$ Images $\{\hat{v}_1, \hat{v}_2, \dots, \hat{v}_{B_O} \}$ from $\mathcal{D}$ \;
    \For{$i=1$ to $B_O$}{
      Select a object or region $r_i$ from image $\hat{v}_i$ \;
      Convert the object class label $\hat{y}_i$ to text description $\hat{t}_i$\;
      Convert the bounding box annotation of $r_i$ to patch annotations $ Y^i = \{y^i_1,y^i_2, \dots, y^i_n\}$ \;
     
    }
    Run forward of $M$ on the mini-batch of image-text pairs $\{\{\hat{v}_1, \hat{t}_1\}, \{\hat{v}_2, \hat{t}_2\}, \dots, \{\hat{v}_{B_O}, \hat{t}_{B_O}\} \}$ and $\{Y^1, Y^2, \dots, Y^{B_O}\}$ to obtain the loss $\mathcal{L}_{PTA}$ \;

    Randomly sample a mini-batch of $B$ Image-Text Pairs $\{\{v_1,t_1\}, \{v_2,t_3\}, \ldots, \{v_{B_D},t_{B_D}\}\}$ from $\mathcal{D}$ \;

    Run forward of $M$ on the mini-batch of image-text pairs $\{\{v_1,t_1\}, \{v_2,t_3\}, \ldots, \{v_{B_D},t_{B_D}\}\}$  to obtain the losses $\mathcal{L}_{ITC}$, $\mathcal{L}_{ITM}$, $\mathcal{L}_{MLM}$, $\mathcal{L}_{Prefix}$ \;
    
    Calculate the overall loss:
    
    $\mathbf{L} = \mathbf{L}_{ITC}+\mathbf{L}_{ITM} +\mathbf{L}_{MLM} + \mathbf{L}_{Prefix} + \mathbf{L}_{PTA}$\;
    
    Backward the overall loss $\mathbf{L}$ and update the parameters of $M$ using gradient descent with learning rate $\alpha$ and the average loss $\mathbf{L}$ over the mini-batch:

    $\theta \leftarrow \theta - \alpha \frac{1}{B} \sum_{i=1}^{B} \nabla_{\theta} \mathcal{L}(\theta; s_i)$ \;

  }

  \Return $M$ with pre-trained parameters $\theta$ \;
 
\end{algorithm}

\vspace{-2ex}
\section{Experiments}
\subsection{Data \& Setup}
\label{data_setup}
Following the previous work~\cite{Li2021AlignBF}, we use the same pre-training dataset with 4M images with texts, which includes two in-domain datasets (MS COCO ~\cite{Lin2014MicrosoftCC} and Visual Genome ~\cite{Krishna2016VisualGC}), and
three web out-of-domain datasets (Conceptual Captions ~\cite{cc}, SBU Captions ~\cite{Ordonez2011Im2TextDI}. See Appendix \ref{sup: data} for more details on the pre-training datasets.

We pre-train the model for 30 epochs with a total batch size of 1024 on 16 NVIDIA A100 GPUs. We use a 6-layer Transformer for both the text encoder and the cross-modal skip-connected network, and a 12-layer Transformer for the decoder. The text encoder is initialized using the first 6 layers of the BERT$_{base}$~\cite{Devlin2019BERTPO} model and the skip-connected network is initialized using the last 6 layers of the BERT$_{base}$. Please see Appendix \ref{sec:pre-training} for more details of the pre-training setting of our model.
 
\begin{table*}[t]
\setlength\tabcolsep{6.6pt}
\centering
\small
\begin{tabular}{l|c|cc|cccccccc|cc}
\toprule[1.0pt]
\multicolumn{1}{c|}{\multirow{3}{*}{Models}}      &
\multicolumn{1}{c|}{\# Pretrain} &
\multicolumn{2}{c|}{VQA} &
\multicolumn{8}{c|}{COCO Caption} & \multicolumn{2}{c}{\multirow{1}{*}{NoCaps}}  \\
\multicolumn{1}{c|}{\multirow{2}{*}{}}      &
\multicolumn{1}{c|}{\multirow{2}{*}{Data}} & \multicolumn{2}{c|}{} &
\multicolumn{4}{c}{Cross-entropy Optimization} & \multicolumn{4}{c|}{CIDEr Optimization} & \multicolumn{2}{c}{}  \\
      &   & Test-dev  & Test-std & B@4 & M & C & S & B@4 & M & C & S & C & S     \\
      
\midrule      
E2E-VLP~\cite{Xu2021E2EVLPEV} & 4M & 73.25  & 73.67 & 36.2 &-&117.3&-&  - & - & - & - & - & - \\
OSCAR~\cite{Li2020OscarOA} & 6.5M & 73.16 & 73.44 & - & - & - & - & 41.7 & 30.6 & 140.0 & 24.5 & 83.4 & 11.4 \\
VinVL~\cite{2021VinVL} & 5.65M & 76.52 & 76.60 & 38.5 & 30.4 & 130.8 & 23.4 & 41.0 & 31.1 & 140.9 & 25.2 & 97.3 & 13.8 \\

METER~\cite{dou2021empirical} & 4M & 77.68 & 77.64  & - & - & - & - & - & - & - & - & - & - \\
BLIP~\cite{li2022blip}  & 14M & 77.54 & 77.62 & 38.6 & - & 129.7 & - & - & - & - & - & \textbf{105.1} & \textbf{14.4}  \\
VLMo~\cite{Wang2021VLMoUV} & 4M & 76.64 & 76.89  & - & - & - & - & - & - & - & - & - & - \\
ViLBERT~\cite{Lu2019ViLBERTPT}  & 3.3M  & 70.63 & 70.92 &  - & - & - & - & - & - & - & - & - & - \\
VisualBERT~\cite{Li2019VisualBERTAS} & 180K & 70.80 & 71.00&- & - & - & - & - & - & - & - & - & -  \\
SimVLM~\cite{Wang2021SimVLMSV} & 1.8B & 77.87 & 78.14  & 39.0 & \textbf{32.9} & \textbf{134.8} & 24.0 & - & - & - & - & - & - \\
ALBEF~\cite{Li2021AlignBF} & 14M & 75.84 & 76.04  & - & - & - & - & - & - & - & - & - & - \\
TRIPS~\cite{Jiang2022TRIPSEV} & 4M &76.23  &76.48 & - & - & - & - & - & - & - & - & - & - \\
mPLUG ~\cite{li2022mplug} & 4M & 77.55  & 77.73 & 39.3 & 30.1 & 132.4 & 23.34 & 41.2 & 30.8 & 140.2 & 25.2  & 98.3 & 12.9 \\
\midrule

\modelname & 4M & \textbf{77.84}  & \textbf{77.91} & \textbf{39.5} & 32.8 & 133.8 &\textbf{24.12} & \textbf{41.5} & \textbf{31.0} & \textbf{140.4} & \textbf{25.1}  &98.9 & 13.1  \\
\bottomrule[1.0pt]
\end{tabular}
\caption{Evaluation Results on VQA, COCO Caption "Karpathy" test split and NoCaps validation set. B@4: BLEU@4, M: METEOR, C: CIDEr, S: SPICE. More details about comparison models in Appendix \ref{sup:comparison models} . } 
\label{table:vqa_caption}
\vspace{-4ex}
\end{table*}

\begin{table*}[t]
\setlength\tabcolsep{7.0pt}
\centering
\small
\begin{tabular}{l|c|cccccc|cccccc}
\toprule[1.0pt]

\multicolumn{1}{c|}{\multirow{2}{*}{Models}}      &
\multicolumn{1}{c|}{\# Pretrain} &
\multicolumn{6}{c|}{MSCOCO (5K test set)} & \multicolumn{6}{c}{Flickr30K (1K test set)} \\
      &  data & \multicolumn{3}{c}{TR} & \multicolumn{3}{c|}{IR} & \multicolumn{3}{c}{TR} & \multicolumn{3}{c}{IR}          \\
\midrule
&&R@1&R@5&R@10&R@1&R@5&R@10&R@1&R@5&R@10&R@1&R@5&R@10 \\ \hline
ALIGN~\cite{jia2021scaling} & 1.8B  & 77.0&93.5&96.9&59.9&83.3&89.8&95.3& 99.8&100.0&84.9&97.4&98.6   \\
OSCAR~\cite{Li2020OscarOA} & 4M  & 70.0&91.1&95.5&54.0&80.8&88.5&-& -&-&-&-&-   \\
E2E-VLP~\cite{Xu2021E2EVLPEV} & 4M     &-& -&-&-&-&- & 86.2 &97.5 &98.92&73.6 & 92.4 &96.0 \\
UNITER~\cite{Chen2020UNITERUI} & 4M     & 65.7&88.6&93.8&52.9&79.9&88.0&87.3& 98.0&99.2&75.6&94.1&96.8  \\
VLMo~\cite{Wang2021VLMoUV} & 4M & 78.2& 94.4& 97.4& 60.6& 84.4& 91.0& 95.3& \textbf{99.9}& 100.0& 84.5& 97.3& 98.6 \\
ALBEF~\cite{Li2021AlignBF} & 14M & 77.6&94.3&97.2&60.7&84.3&90.5&95.9& 99.8&100.0&85.6&97.5& \textbf{98.9}                 \\
BLIP~\cite{li2022blip} & 14M & 80.6 &95.2&97.6&63.1&85.3&91.1&96.6& 99.8&100.0&87.2&97.5&98.8                 \\
TRIPS~\cite{Jiang2022TRIPSEV} & 14M & 78.1 &94.8 &97.6 & 61.3& 84.3& 91.4&96.3& 99.8& 100.0 &85.8 &98.1& 99.0 \\
mPLUG~\cite{li2022mplug} & 4M  & 80.2 &95.1&97.7&62.5&84.8&90.9 &96.4& 99.8&100.0& 86.5&97.5&98.8   \\
\midrule
\modelname & 4M  & \textbf{80.8} & \textbf{95.6} & \textbf{98.1} & \textbf{63.6} & \textbf{85.6} & \textbf{91.6}&\textbf{96.8}& 99.8&\textbf{100.0}&\textbf{87.3} & \textbf{97.9} & \textbf{98.9}   \\
\bottomrule[1.0pt]
\end{tabular}          \\
\caption{Image-text retrieval results on Flickr30K and COCO datasets.}
\label{table:retrieval}
\vspace{-5ex}
\end{table*}

\subsection{Main Result}

We evaluate our model\modelname on four widely explored vision-language downstream tasks: Visual Question Answering (VQA), Cross-modal Retrieval, Image Caption, and Visual Grounding (VG). We plug the Text-aware Patch Detector before the 6-th Transformer layer in the ViT encoder and set the keeping ratio to 50\%, achieving the desired trade-off between the downstream task performance and the model inference speed. The fine-tuning hyperparameters are described in Appendix \ref{sec:downsteam}. Details of the comparison methods are in Appendix \ref{sup:comparison models}.

\subsubsection{Visual Question Answering}

The VQA task \cite{Agrawal2015VQAVQ} requires the model to answer natural language questions given an image.  During fine-tuning and inference of VQA, we feed the [CLS] token of the question to the TPD to detect the question-relevant patch tokens (We also give the visualization of the detected tokens in Figure~\ref{vqacase1} which indicates the effectiveness and generalization of TPD). We follow \cite{Li2021AlignBF} and consider VQA as an answer-generation problem. We report test-dev and test-std scores by submitting our results to the evaluation server\footnote{https://eval.ai/web/challenges/challenge-page/830/overview} in Table \ref{table:vqa_caption}. Compared with the VLP baselines, our \modelname can get the better performance (e.g. 77.84 on VQA test-dev) with SOTAs under the same image resolution (384 $\times$ 384) and even speed up about 88\% of model inference(see the report results in Table~\ref{table_efficient} and Table~\ref{table: image_size}). Furthermore, when we increase the image resolution to 512 $\times$ 512 (as shown in Table~\ref{table: image_size}), we can achieve better performance (e.g. 78.25 on VQA test-dev. ) while keeping a similar inference computation cost with the baselines mPLUG~\cite{li2022mplug} (e.g. 65.23 of \modelname$_{512 \times 512}$ VS 63.57 of mPLUG~\cite{li2022mplug} on Throughput.). The results demonstrate the effectiveness and efficiency of \modelname.

\subsubsection{Image Captioning} 

As there is no textual input in the image caption task, we directly detect the patches based on the vision information where we use the attention weight of the image [CLS] token to other image tokens as the detection scores and fusion the image tokens with low attention weight of image [CLS] token. Following~\cite{Li2020OscarOA}, we first fine-tune \modelname with cross-entropy loss and then with CIDEr optimization~\cite{scst} for extra 5 epochs. As shown in Table~\ref{table:vqa_caption}, \modelname can get comparable or better results with SOTA models on both COCO Caption~\cite{Lin2014MicrosoftCC} and Nocaps~\cite{nocaps} datasets.
\begin{table}[t]
\small
\centering
 \setlength{\tabcolsep}{6.4mm}{
\begin{tabular}{@{}lccccc@{}}
\toprule[1.0pt]
\multicolumn{1}{c}{\multirow{2}{*}{Model}} & \multicolumn{3}{c}{RefCOCO+} \\
\multicolumn{1}{c}{}          & val   & testA   & testB          \\ \midrule
UNITER  \cite{Chen2020UNITERUI} & 75.90    & 81.45   & 66.70         \\
VL-BERT \cite{Su2020VLBERTPO} &72.59  & 78.57 & 62.30   \\
ViLBERT\cite{Lu2019ViLBERTPT} & 72.34&  78.52 &   62.61             \\
VILLA \cite{gan2020large}  & 76.17    & 81.54   & 66.84            \\
MDETR \cite{kamath2021mdetr}   & 79.52    & 84.09   & 70.62      \\
UNICORN \cite{DBLP:journals/corr/abs-2111-12085} & 80.30    & 85.05   & \textbf{71.88}      \\
mPLUG  \cite{li2022mplug}   & 80.07  &   85.21 &  71.03       \\ \hline
\modelname     & \textbf{80.37}  &   \textbf{86.03} & 71.81      \\ 
\bottomrule[1.0pt]
\end{tabular}}
\caption{Evaluation results of Visual grounding on ReferCOCO+. We use the accuracy of IOU 0.5 on visual grounding (a prediction is right if the IoU between the grounding-truth box and the predicted bounding box is larger than 0.5)}
\label{tab:visual_grounding}
 \vspace{-4ex}
\end{table}

\begin{table}[htbp]
\small
\setlength{\tabcolsep}{3.0mm}
\begin{tabular}{lcccc}
\toprule[1.0pt]
Models & VQA & FLOPs & Throughput & Latency \\ \hline
UNITER \cite{Chen2020UNITERUI}  & 72.70  &    949.90    &     6.42       &     ~870ms    \\
OSCAR \cite{Li2020OscarOA}  & 73.16   &    956.40  &    6.35      &     ~860ms    \\
VinVL \cite{2021VinVL}  &  76.52   &    1023.30    &    7.32         &   ~640ms     \\
E2E-VLP \cite{Xu2021E2EVLPEV} & 73.25  &    144.3     &     80.23     &     ~70ms    \\
ViLT \cite{Kim2021ViLTVT}   &  71.26 &    55.40   & 247.530       &  ~19ms    \\ 
ALBEF \cite{Li2021AlignBF}  &  74.54  &33.42    &     197.52    &      ~22ms \\
TRIPS \cite{Jiang2022TRIPSEV}& 76.23 &        20.89          &343.05   & 11ms \\
mPLUG \cite{li2022mplug}  & 77.55  & 36.63    &     186.42    &      ~24ms   \\\hline
\modelname & \textbf{77.84} &    \textbf{19.84}  &     \textbf{349.71}       &   \textbf{~10ms}      \\  \bottomrule[1.0pt]
\end{tabular}
\centering
\caption{The comparison of the efficiency of different models. Here, we report the VQA test-dev result and FLOPs, throughput, and latency. The FLOPs results of the baselines come from  \cite{Kim2021ViLTVT}. Since FLOPs are proportional to input size, for a fair comparison, we keep same the input size with \cite{Kim2021ViLTVT}, which is 197 for image patches (the image resolution is $224 \times 224$ ) length and 40 for text tokens length. We keep the same setting when calculating throughput and latency.}
\label{table_efficient}
\vspace{-6ex}
\end{table}
 
\subsubsection{Image-Text Retrieval}
We conduct experiments for both image-to-text retrieval (TR) and text-to-image retrieval (IR) on MSCOCO \cite{Lin2014MicrosoftCC} and Flickr30K \cite{Plummer2015Flickr30kEC} datasets. We jointly optimize the ITC loss and the ITM loss during fine-tuning. 
The results are reported in Table \ref{table:retrieval}. As shown in Table \ref{table:retrieval}, the experimental results show that our model gets comparable performance with other VLP baselines. For more details, please refer to the Appendix~\ref{sec:downsteam}
 
\subsubsection{Visual Grounding} 
Following the setting of mPLUG~\cite{li2022mplug}, we also evaluate \modelname on the visual grounding task, which requires models to localize the referred object in the image based on a given text query. In this task, we feed the text query to the TPD to detect the query-relevant patch tokens, then, instead of directly regressing the bounding boxes, we concatenate detected patch features and attended textual features and feed them into the multi-modal decoder to predict the coordinates. Table \ref{tab:visual_grounding} demonstrates the performance of \modelname in the visual grounding task. \modelname achieves comparable results with competitive baseline methods.


\begin{table*}[h]
\centering
\small
\setlength{\tabcolsep}{7.6mm}{
\begin{tabular}{@{}ccccccc@{}}
\toprule[1.0pt]
Detection Location         & Keeping ratio     & image size     & VQA test-dev& FLOPs(G) & Throughput \\ \midrule
 - & - & $384 \times 384$ &  77.55  & 84.872  & 63.57 \\ \hline
 {[}6{]}            & 50\% & $224\times224$ &     76.83        &\textbf{19.84} &  \textbf{ 349.71}   \\
 {[}6{]}            & 50\% & $256\times256$ &      77.11       &    23.04 &  303.03     \\
 {[}6{]}            & 50\% & $304\times304$ &    77.32         &   32.97  &  247.62\\
 {[}6{]}            & 50\% & $384\times384$ &      77.84       &    47.56 &  144.38   \\
 {[}6{]}            & 50\% & $464\times464$ &     78.03        &  75.99  &  81.02  \\
 {[}6{]}            & 50\% & $512\times512$ &     \textbf{78.25}    &  83.24 &  65.23   \\\bottomrule[1.0pt]
\end{tabular}}
\centering

\caption{Results of \modelname finetuning on VQA task with different resolution images. The Settings for calculating FLOPs and throughput are the same as Table \ref{table_efficient} except for the image resolution. The first row in the table reports the result of our baseline mPLUG}
\label{table: image_size}
\vspace{-5ex}
\end{table*}
\subsection{Efficiency of Text-aware Patch Detection}
 
\begin{figure}[htbp]
\centering
\subfigure[Keeping Ratio - GPU MEM]{
\centering
\includegraphics[width=1.5in]{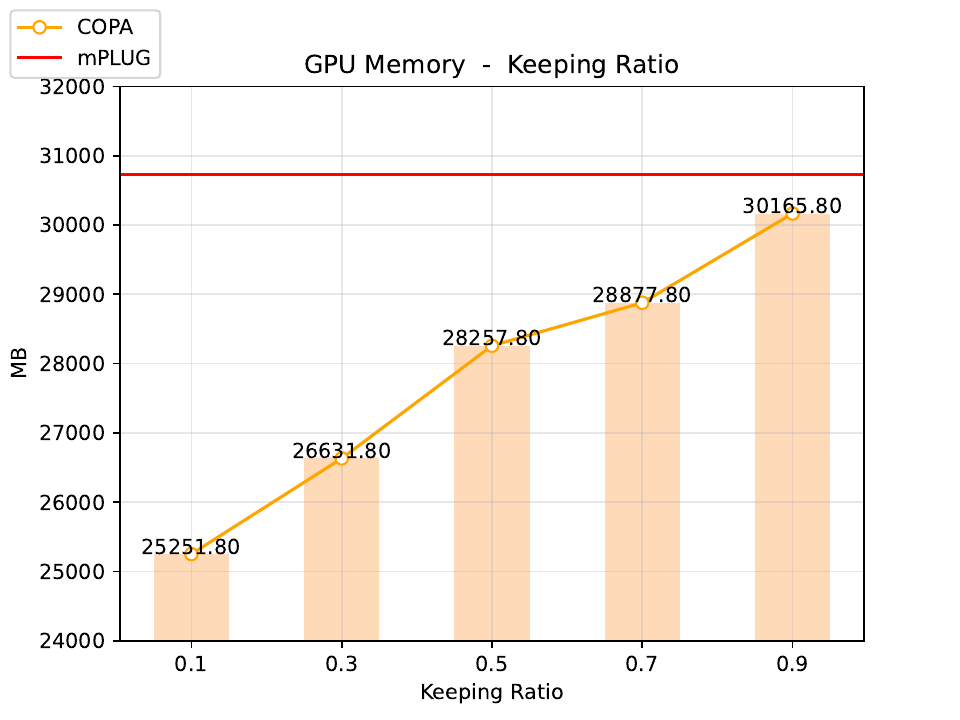}
}%
\subfigure[Detection Location -  GPU MEM]{
\centering
\includegraphics[width=1.5in]{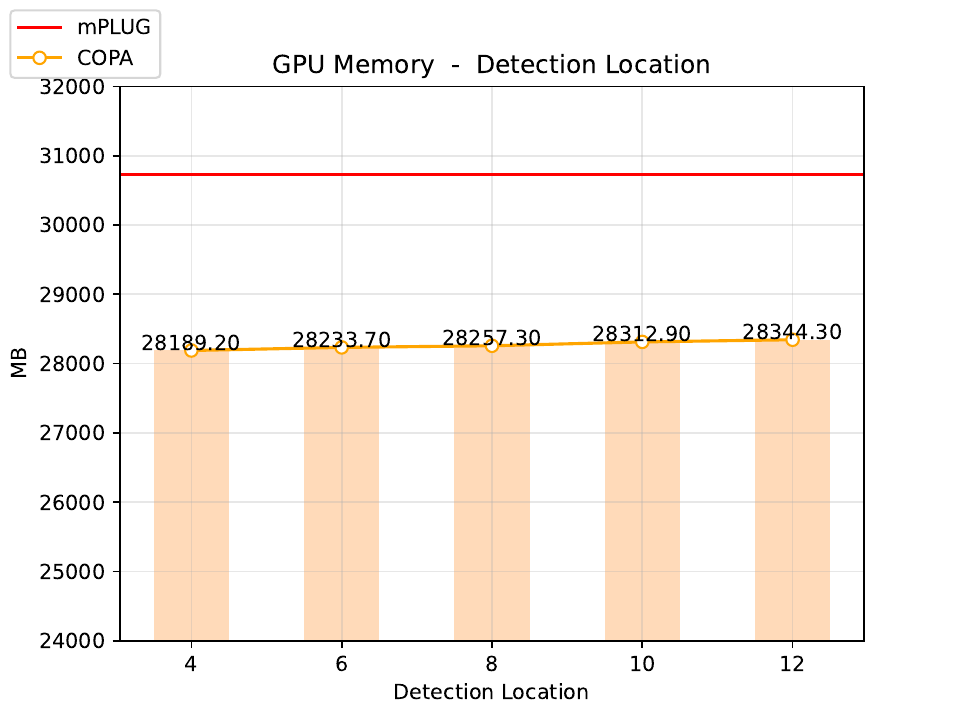}
}%
\caption{The single GPU Memory cost of pre-training of \modelname with different keeping ratios and detection locations of TPD. We set the batch size to 512, the image size to 256 and the text length to 25. The red line in sub-figure (a) is the GPU memory cost of the baseline model mPLUG \cite{li2022mplug}}
\label{fig:gpu training time}
 \vspace{-2ex}
\end{figure}


To assess the efficiency of the Text-aware Patch Detection mechanism, we first compare the computational complexity of various models. We report the Floating Point Operations Per Second (FLOPs), a widely used evaluation metric for model computational complexity. Additionally, to evaluate the computational speed of our model, we compare the throughput and latency of different models. We use a Xeon Platinum 8163 CPU and an NVIDIA V100 GPU to measure latency and throughput.

As illustrated in Table \ref{table_efficient}, \modelname exhibits not only the lowest computational complexity (e.g., 19.84 FLOPs) but also the fastest computational speed (e.g., 349.71 throughput and 13ms latency). Moreover, as shown in Figure \ref{fig:gpu training time}, we evaluate the single GPU memory cost under varying keeping ratios and detection location, which refers to the position where TPD is integrated into the ViT backbone. For example, Detection Location=6 indicates that TPD is inserted before the 6th transformer layer in the ViT backbone. The results demonstrate that Text-aware Patch Detection significantly reduces GPU memory usage compared to the baseline model. We also observe that the keeping ratio influences GPU memory consumption, while the detection location of TPD does not impact it.

\subsection{ The Impact of Detection Location and Keeping Ratio }

To investigate the influence of detection location in the ViT backbone and keeping ratio on the efficiency and effectiveness of \modelname, we train \modelname using different detection locations and keeping ratios. Note that when calculating FLOPs and throughput, we set the input image size to $224 \times 224$ and input text length to 40. As depicted in Table~\ref{table:impactofKR}, two main conclusions can be drawn:

First, incorporating the Text-aware Patch Detector (TPD) before shallower layers can reduce computational complexity but at the cost of accuracy. For instance, when TPD is placed before the 4th layer, there is a significant increase in throughput, but the accuracy drops considerably. A possible explanation is that the patch embedding in shallow layers may not adequately represent visual semantics, making it challenging to learn the fine-grained patch-text alignment and subsequently leading to a decline in accuracy.

Second, introducing too many undetected image tokens into the TPD module can significantly impair downstream task performance. For example, when positioning the TPD module before the 4th layer in ViT and setting the keeping ratio to 10\%, the performance on the VQA task decreases to 76.48, compared to 77.84 achieved by the model with a 50\% keeping ratio in the 6th layer.


                        


\subsection{Finetuning on Higher Resolution Images}
\label{HigherImages}
 
We can regulate the computational cost by fusing different numbers of inattentive tokens. To this end, we fine-tune \modelname on the VQA task, using images with varying resolutions as input. The results are reported in Table \ref{table: image_size}. The experimental findings indicate that by increasing the input image resolution, the model benefits from processing more image tokens, resulting in improved performance. For instance, by fine-tuning \modelname with 512$\times$512 resolution images, we can achieve a score of 78.25 on VQA, surpassing the baseline fine-tuned with 384$\times$384 images while maintaining a similar computational complexity.


\subsection{Ablation Study}
 \subsubsection{Effectiveness of Text-aware Patch Detector}
 We also perform ablation studies to investigate the effects of our proposed pre-training task Patch-Text Alignment (PTA) and Text-aware Patch Detector (TPD). In Table \ref{table6}, w/o PTA indicates we remove the PTA task but keep the TPD in the visual backbone. However, without the PTA task, the TPD can not be optimized and thus is ineffective, therefore, we replace the TPD with a simple strategy in which we directly detect the patch in a transformer layer based on the self-attention weights of image [CLS] token to other patch tokens. As shown in Table \ref{table6}, we find that without the text guidance,  w/o PTA, which directly detects patches based on visual information, will get a significant accuracy drop on VQA compared with the baseline model mPLUG. On the contrary, \modelname detects the text-relevant patches based on the TPD and can even get a slight improvement compared with the baseline model mPLUG, which indicates the effectiveness of TPD and PTA. 
 
 \subsubsection{Effectiveness of Patch-Text Alignment}
 w/o TPD means we remove the TPD but keep the PTA pre-training task, compared with the baseline model mPLUG~\cite{li2022mplug} (w/o TPD \& w/o TPA), we find that even though the efficiency of the model can not be improved compared with the baseline, we can get a remarkable improvement on VQA. This experiment results not only shows the efficiency of the text-aware Patch Detection mechanism but also indicate the effectiveness of PTA, which enables our model to learn fine-grained cross-modal alignment, thus leading to improvement in VQA task.





\subsection{Extension to Single-stream Model }
\begin{table}[htbp]
\setlength{\tabcolsep}{2.2mm}
\small
\begin{tabular}{ccccccccc}
\toprule[1.0pt]
\multirow{2}{*}{Models} & \multirow{2}{*}{EL} &\multirow{2}{*}{KR} &\multirow{2}{*}{FLOPs(G)}&\multirow{2}{*}{Throughput} & \multicolumn{1}{c}{VQA}  \\
                        &               &  &  &            & Test-dev     \\ \hline
                        
\modelname-S        & 6 & 50\%        &    26.23   &    \textbf{436.74}      & 71.51         \\
 ViLT     & - &     -     &   55.40        &    247.53     & 71.26     \\ \hline
\modelname       &6 & 50\%     & \textbf{19.84}    &  349.71      &  \textbf{77.84}               \\
 mPLUG  & - &    -      &  36.63      &  186.42  &  77.55      \\ \bottomrule[1.0pt]
\end{tabular}
\centering
\caption{The evaluate results of \modelname, \modelname-S and their baselines ViLT and mPLUG on VQA test-dev.  The setting for calculating FLOPs and throughput is the same as Table ~\ref{table_efficient}. KR refers to Keeping Ratio, EL refers to Detection Location.}
\label{table3}
\vspace{-5ex}
\end{table}

The proposed Text-aware Patch Detection mechanism can also be extended to single-stream models by incorporating the TPD into the multimodal encoder and pre-training the model with the \PretrainTaskName task. To verify the effectiveness of TPD and \PretrainTaskName task in single-stream models, we first implement a single-stream model (\modelname-S) based on the ViLT \cite{Kim2021ViLTVT} framework, which employs a visual transformer as the cross-modal encoder. Next, we insert the TPD before the 6th layer of the cross-modal encoder and pre-train it with the PTA task (For the single-stream model, the TPD can be directly integrated into the cross-modal encoder and detect patches based on the overall [CLS] token's attention value.). We then evaluate the downstream task performance, computational complexity, and inference speed of \modelname and \modelname-S (both with and without Text-aware Patch Detection). The results are shown in Table \ref{table3}, and we observe consistent improvements in inference speed and downstream task performance for both \modelname and \modelname-S when incorporating Text-aware Patch Detection and Patch-Text Alignment. These results indicate that the proposed image patch selection mechanism is not only efficient but also effective. Notably, compared to the dual-stream model \modelname, \modelname-S has faster inference due to the parameter efficiency of the single-stream model. However, its performance falls short of state-of-the-art performance on downstream VL tasks.

\subsection{Case Study}
\label{sec:case_vqa}

 \begin{figure}[!htbp]
\centering
\includegraphics[width=0.98\linewidth]{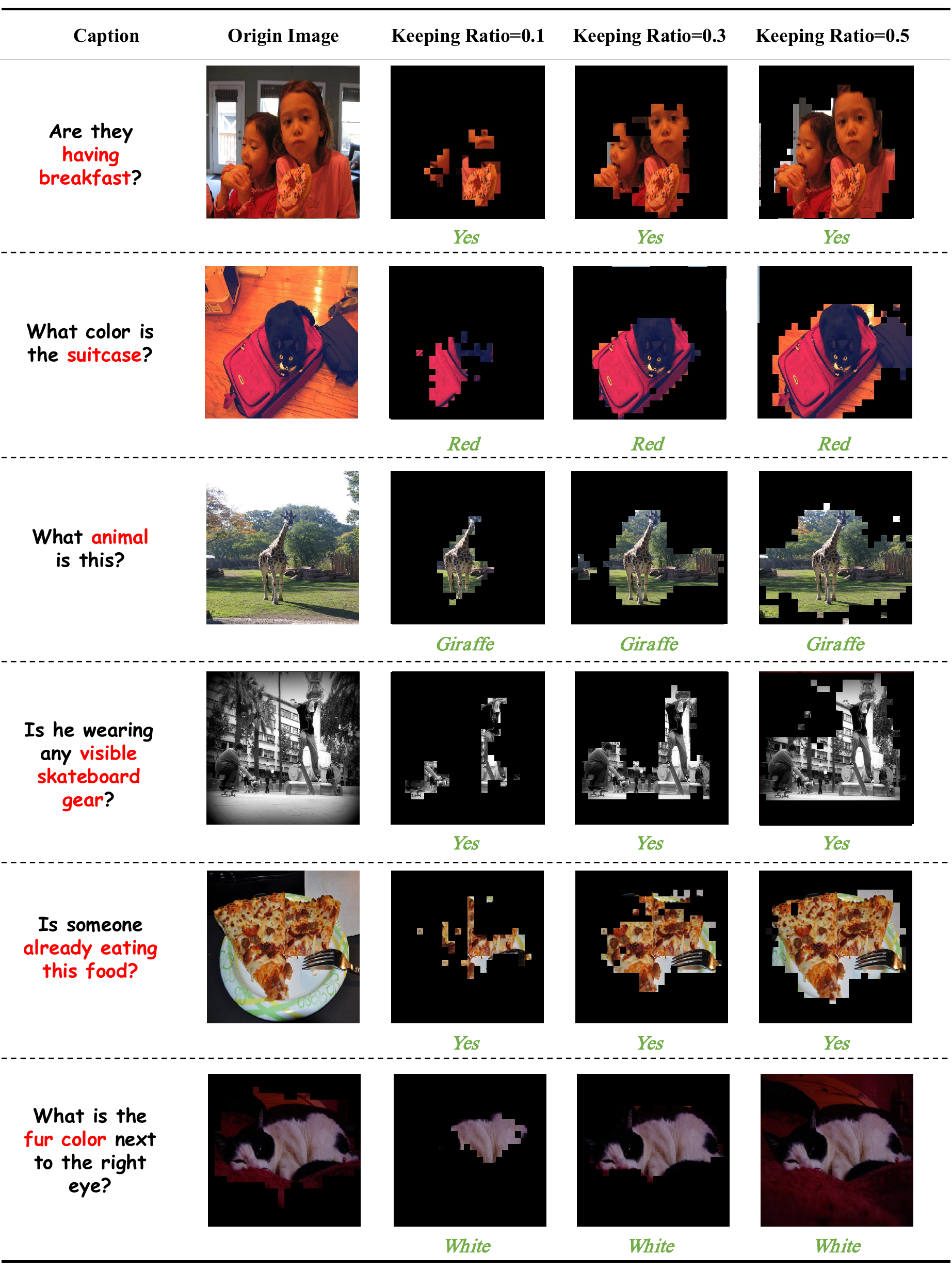} 
\caption{The visualization of the VQA case and the detected text-relevant image patches. We set the detection location to 6.}
\vspace{-2ex}
\label{vqacase1}

\end{figure}
The proposed Text-aware Patch Detector (TPD) identifies text-consistent image tokens in the vision backbone and retains the detected image patches. To further investigate the effectiveness of the TPD, we visualize the VQA case and the detected text-relevant image patches in Figure \ref{vqacase1}. It is evident that based on the text questions, the TPD module can effectively detect relevant patches, and even when preserving only 10\% of the detected patches, our model can still produce correct answers. For instance, in the first case, the question is "Are they having breakfast?", and the TPD effectively detects patches of food and the girl's mouth in the image, which are highly relevant to the question. In the second case, the question is "What color is the suitcase?", indicating that the red suitcase in the image is text-relevant, while other visual objects like the black cat are text-irrelevant. As illustrated in Figure \ref{vqacase1}, the TPD effectively detects text-relevant patches, which help our model predict the correct answer. It is worth noting that these examples are not cherry-picked, and this phenomenon is commonly observed in other samples.

\section{Conclusion}
We have presented \modelname, an efficient and effective VLP model that successfully learns fine-grained patch-text alignment and reduces lengthy visual sequences for streamlined training. Specifically, we devise a \PretrainTaskName pre-training task (PTA) based on a Text-aware Patch Detector (TPD). The TPD is incorporated into the ViT backbone to identify text-relevant patches and eliminate redundant ones. PTA allows our model to learn fine-grained patch-text alignment end-to-end by jointly optimizing with other pre-training tasks. Experiments demonstrate that our method enhances efficiency through the reduction of visual sequences while maintaining or even improving the performance of downstream tasks.

\bibliographystyle{ACM-Reference-Format}
\balance
\bibliography{release}

\clearpage

\appendix

\section{Datasets}
\label{sup: data}
Following the previous work~\cite{Li2021AlignBF}, we use the same pre-training dataset with 4M images with texts, which includes two in-domain datasets (MS COCO ~\cite{Lin2014MicrosoftCC} and Visual Genome ~\cite{Krishna2016VisualGC}), and
three web out-domain datasets (Conceptual Captions~\cite{cc}, SBU Captions ~\cite{Ordonez2011Im2TextDI}. 

\begin{table}[htbp]
\setlength\tabcolsep{12pt}
\centering

\begin{tabular}{l|cccc}
\toprule[1.5pt]
  &  COCO & VG & SBU & CC3M \\
\midrule
image & 113K & 100K & 860K & 3M   \\
text & 567K & 769K & 860K & 3M \\
\bottomrule[1.5pt]
\end{tabular} 
\caption{Statistics of the pre-training datasets.}
\label{table:pretraindata}
\vspace{-5ex}
\end{table}

\begin{table}[htbp]
\setlength\tabcolsep{10.3pt}
\centering

\begin{tabular}{l|cccc}
\toprule[1.5pt]
     &  image & Captions & Objects & Regions \\
\midrule
COCO & 0.11M & 0.55M & 0.45M & -   \\
VG   & 0.10M & - & 2.0M & 3.7M \\
\bottomrule[1.5pt]
\end{tabular} 
\caption{Statistics of objects/regions annotations used in the pre-training.}
\label{table:objectdata}
\vspace{-5ex}
\end{table}

Table \ref{table:pretraindata} shows the statistics of the 4M images with texts used in the pre-training stage. Besides, As shown in table~\ref{table:objectdata} we use also use the objects/regions annotations from COCO\cite{Lin2014MicrosoftCC} and VG \cite{Krishna2016VisualGC} datasets, and we give statistics of object and region annotations of each dataset. Note that we use the object/region annotations provided by ~\citet{Zeng2021xvlm} thus we follow their setting which filtered out some samples because of: 1) invalid annotations (e.g. negative values for bounding boxes or boxes being outside of the images); 2) boxes being too small (< 1\%); 3) highly overlapped textual descriptions of regions (>75\%), etc. After pre-processing,  we keep COCO objects 446,873 (from 859,999), VG objects 2,043,927 (from 3,802,349), VG regions 3,699,598 (from 5,402,953).

\section{Pre-training Objectives}
\label{pretraing objectives}

We pre-train our model with five standard objectives: \PretrainTaskName (PTA), Image-Text Contrastive learning (ITC), Image-Text Matching (ITM), Masked Language Modeling (MLM), Prefix Language Modeling (PrefixLM). These pre-training tasks are optimized jointly. As we have talked about the \PretrainTaskName before, in this subsection, we only introduce the last four pre-training tasks.

\noindent \textbf{Image-text Contrastive (ITC)} For \modelname, We follow the \cite{Li2021AlignBF} and apply ITC to align the image representation and text representation from the unimodal encoders. For the image, the image feature corresponding to the image [CLS] token is chosen as the image representation. For the text, the text token feature corresponding to the text [CLS] token is the text representation.

\noindent \textbf{Image-Text Matching (ITM)}  The goal of image-text matching is to predict whether the input image and text are matched.  We follow the design of \cite{Li2021AlignBF} and select hard negative image-text pairs based on the contrastive text-image similarity. We take the text [CLS] embedding of the multimodal encoder's output as the joint representation, followed by a Multi-Layer Perceptron (MLP) layer for prediction.

\noindent \textbf{Masked Language Modeling (MLM)} The task setup is basically the same as in BERT~\cite{Devlin2019BERTPO}, where we randomly mask 15$\%$ of tokens in text, and the model is asked to predict these masked words with the cross-modal representations.

\noindent \textbf{Prefix Language Modeling (PrefixLM).} This task aims to generate the caption given an image and predict the text segment subsequent to the cross-modal context as ~\cite{bi2020palm}. It optimizes a cross-entropy loss by maximizing the likelihood of text in an autoregressive manner.

\section{More Experiment Results.}
\subsection{Generalization of Text-aware Patch Detector}
 
\begin{figure}[h]
\centering
\subfigure[ ACC of TPD]{
\centering
\includegraphics[width=1.6in]{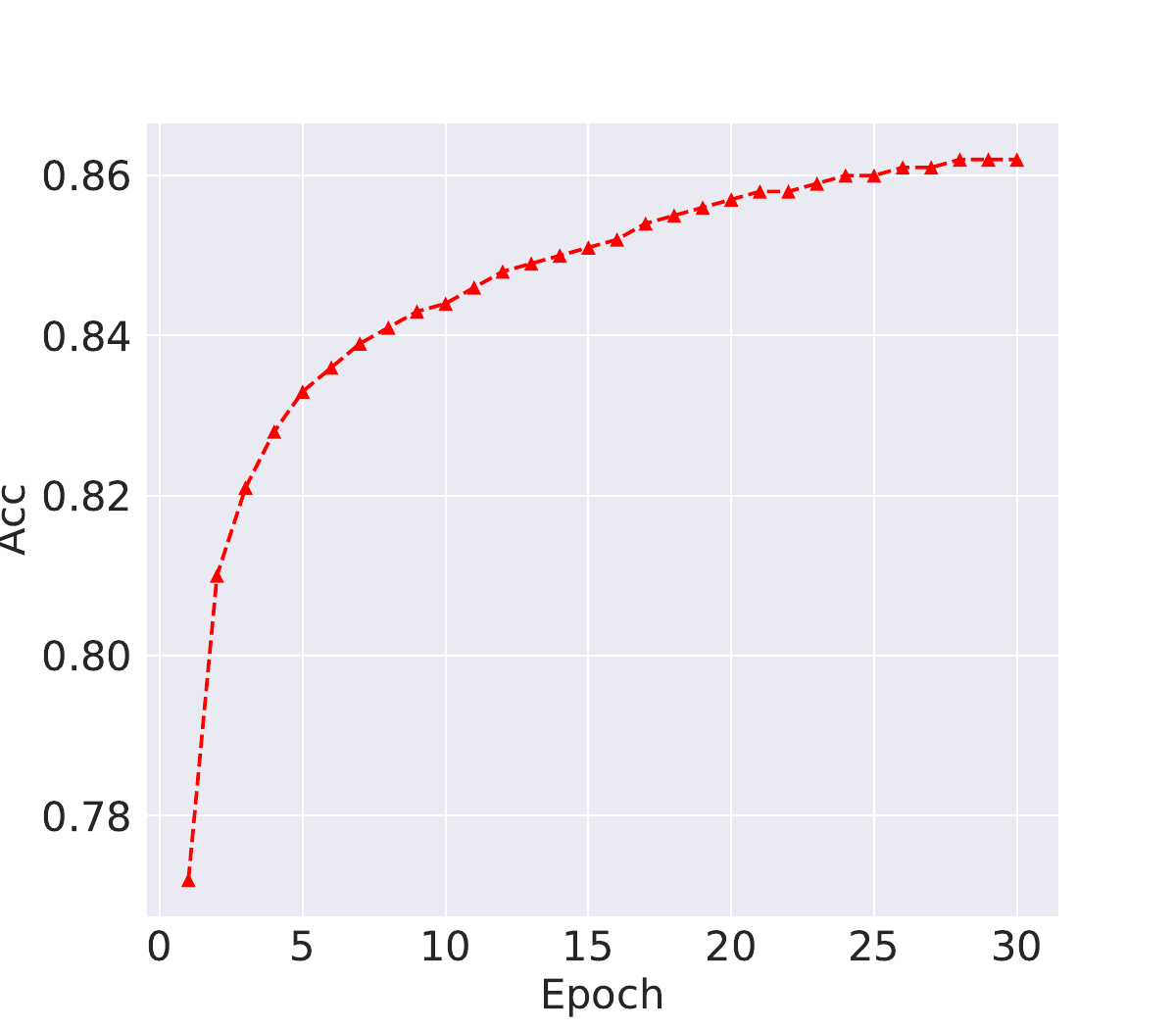}
}%
\subfigure[  Recall of TPD]{
\centering
\includegraphics[width=1.6in]{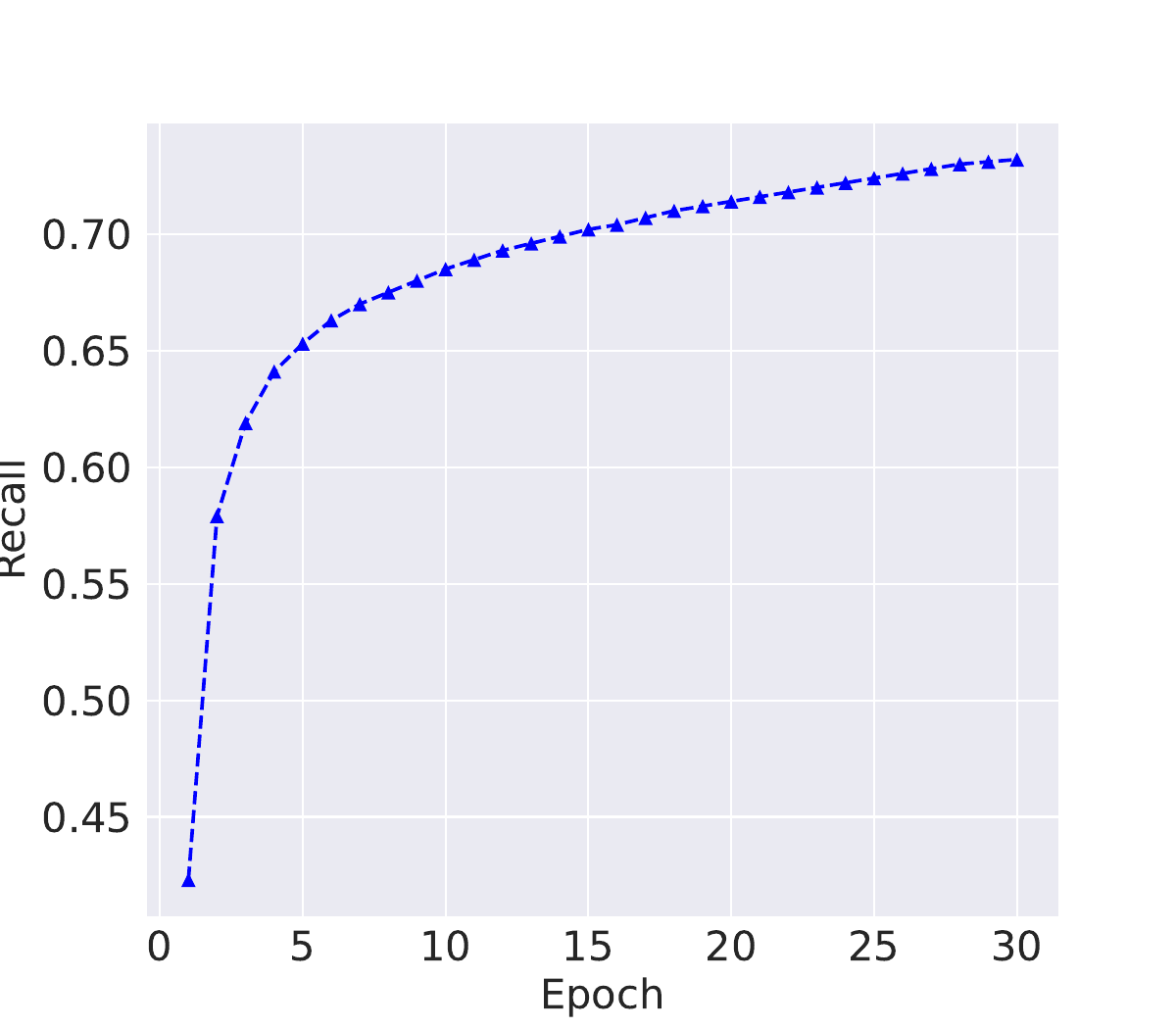}
}%
\caption{The visualization of Accuracy and Recall of TPD on the 10K test dataset randomly sampled from CC~\cite{cc}.}
\label{fig:acc_recall_TPD}
 \vspace{-0.2cm}
\end{figure}
\noindent The Text-aware Patch Detector (TPD) is pre-trained based on the \PretrainTaskName task, utilizing object/region annotations from the COCO~\cite{Lin2014MicrosoftCC} and VG~\cite{Krishna2016VisualGC} datasets as supervised signals. Our subsequent experiments demonstrate that using these two relatively small-scale datasets alone is sufficient to train a robust and generalizable TPD without requiring additional object/region annotations while maintaining the scalability of web-based pre-training data.

In detail, we sample 10K image-text pairs from CC~\cite{cc} dataset, which is crawled from the web and potentially contains out-of-domain data. We then employ the state-of-the-art open-set object detection model, Grounding DINO~\cite{Liu2023GroundingDM}, to detect regions in each image corresponding to the text. Following the approach described in subsection \ref{subsection:PTA}, we convert these regions into patch-level labels. We subsequently evaluate the accuracy and recall of TPD in detecting patches on this test dataset.

As depicted in Figure~\ref{fig:acc_recall_TPD}, the accuracy and recall of TPD's patch predictions gradually improve with each training epoch. By the 30th epoch, TPD's accuracy reaches 0.87, and its recall approaches 0.74. These results suggest that TPD is effective and robust in detecting text-related image patches, regardless of whether the image-text pairs originate from manual annotations or are crawled from the internet. This showcases TPD's remarkable generalizability and effectiveness, allowing us to confidently scale up the size of the web-based pre-training dataset without compromising TPD's ability to handle out-of-domain data from the web.

\begin{table}[t]
\small
 \setlength{\tabcolsep}{1.9mm}{
\begin{tabular}{ccccccc}
\toprule[1.0pt]
 Location & Keeping ratio     & VQA & FLOPs (G) & Throughput  \\ \hline
 {[}4{]}   & 10\%      &   76.48    &   10.15  &  514.26 \\
{[}4{]}   & 30\%      &    77.02    &   14.84  &  443.48 \\
{[}4{]}   &50\%       &    77.33    &   16.62  & 363.65  \\ 
 {[}4{]}  & 70\%      &     77.52   &  23.32   &  337.45 \\\hline
 {[}6{]}  & 10\%       &  76.89     &   16.51  & 418.95 \\
{[}6{]}  & 30\%       &  77.53     &   17.84   & 361.05  \\
{[}6{]}  & 50\%      &  77.84      &   19.84   & 349.71\\
{[}6{]}  & 70\%      &  77.89     &   26.77    & 306.11 \\\hline
{[}8{]}  & 10\%       &  76.93     &   22.45   & 275.13  \\
{[}8{]}  & 30\%     &   77.62      &   23.36   & 310.56\\
{[}8{]}  & 50\%     &  77.89     &  25.66      &  273.06 \\
{[}8{]}  & 70\%     &   77.93     & 30.22      & 221.88 \\ \hline
 -       & 100\% &   77.98    &  36.63   & 186.42  \\ \bottomrule[1.0pt]
\end{tabular}}
\centering
\caption{Results of pre-training and finetuning \modelname with different locations and keeping ratios. We report the text-dev score results of VQA, FLOPs, and Throughput. The throughput (image-text/s) is measured on an NVIDIA V100 GPU using the largest possible batch size for our model.  }
\label{table:impactofKR}
\vspace{-5ex}
\end{table}

\begin{table}[t]
\small
\setlength{\tabcolsep}{5.1mm}{
\begin{tabular}{@{}lcccc@{}}
\toprule[1.0pt]
model    &  VQA & FLOPs(G) & Throughput \\ \midrule

\modelname &  77.84   & 19.84   &    349.71       \\
  -w/o PTA &  77.12 &   19.43  &     352.56  \\ 
  -w/o TPD &  77.98 &   36.63  &     186.42    \\ \hline
  mPLUG &  77.55 &   36.63  &     186.42  \\
  \bottomrule[1.0pt]
\end{tabular}}
\caption{The result of ablations. We finetune \modelname on VQA and report test-dev results. The setting for calculating FLOPs and throughput is the same as Table~\ref{table_efficient}. For -w/o PTA, we keep the same setting with \modelname (detection location = 6 and keeping ratio = 50\%).  }
\label{table6}
 \vspace{-5ex}
\end{table}
 
\vspace{-1ex}

\section{Pre-training Details}
\label{sec:pre-training}

We use the AdamW ~\cite{Loshchilov2019DecoupledWD} optimizer with a weight decay of 0.02. The learning rate is warmed-up to 1e-5 (ViT-B/16) and 1e-4 (BERT$_{base}$)  in the first 1000 iterations, and decayed to 1e-6 following a cosine schedule. \modelname is pre-trained for about 20 epochs with 8*A100-80G GPUs on the 4M pre-training dataset for 41 hours. During pre-training, the batch size on a single GPU is 512, and the overall batch size is 1024.

During pre-training, we take random image crops of resolution 256 $\times$ 256 as input and also apply RandAugment ~\cite{cubuk2020randaugment} to improve the generalization of vision encoders. For VQA and image captioning tasks, we increase the image resolution during finetuning. For image-text contrastive learning, the queue size is set as 65,536, and the momentum coefficient is set as 0.995.

\section{Downstream Task Details}
\label{sec:downsteam}

We evaluate \modelname on the four downstream vision-language tasks. The hyperparameters that we use for finetuning on the downstream tasks are listed in Table \ref{table:finetune-hyper}. Following ~\citep{Li2021AlignBF}, all tasks adopt RandAugment, AdamW optimizer with a weight decay of 0.05 and a cosine learning rate schedule. Next, we introduce the dataset settings in detail.


\paragraph{VQA.} The VQA task ~\cite{Agrawal2015VQAVQ} requires the model to answer natural language questions given an image. Most methods~\cite{Tan2019LXMERTLC,Wang2021VLMoUV,Li2020OscarOA,Wang2021SimVLMSV} deal with visual question-answering tasks as multi-label classification on pre-defined answer sets. This strategy achieves strong performance, but it is not suitable for real-world open scenarios. We conduct an experiment on the VQA2.0 dataset ~\citep{goyal2017making}, which contains 83k/41k/81k images for training/validation/test. Following ~\citep{Li2021AlignBF}, we use both training and validation splits for training, and incorporate additional training data from Visual Genome~\citep{Krishna2016VisualGC}. Besides, we concatenate the question with the object labels and OCR tokens extracted from the image.
\begin{table}
\setlength\tabcolsep{2pt}
\centering

\begin{tabular}{l|ccc}
\toprule
Task  &  LR (ViT-L/BERT$_{base}$) & batch size & epochs  \\
\midrule
VQA & 2e-5/5e-6 & 1024 &  8 \\
Captioning$\dagger$ & 1e-5\&8e-7 & 256& 5 \\
Retrieval & 1e-5/2e-6 & 256& 5 \\
Visual Grounding & 2e-5/2e-6 & 512& 120 \\
\bottomrule
\end{tabular} 
\caption{Finetuning hyperparameters for downstream tasks. $\dagger$ denotes two-stage fine-tuning.}
\label{table:finetune-hyper}
\vspace{-5ex}
\end{table}
\paragraph{Image Captioning.} The image captioning task requires a model to generate an appropriate and fluent caption for a given image. We evaluate image captioning on two datasets COCO Caption~\cite{Lin2014MicrosoftCC} and NoCaps~\cite{nocaps}. \modelname finetuned with training data of COCO Caption is tested on both of the datasets. We train \modelname on the MS COCO Caption and test on the same Karpathy split~\cite{Li2020OscarOA,Wang2021SimVLMSV} and NoCaps validation set. Following~\cite{Li2020OscarOA}, we first fine-tune \modelname with the cross-entropy loss for 5 epochs with a learning rate of 1e-5 and a batch size of 256. Based on the fine-tuned model, we then fine-tune it with CIDEr optimization~\cite{scst} for extra 5 epochs with a smaller learning rate of 8e-7.  We use the best checkpoint on COCO Caption and predict on the Nocaps validation set directly. During inference, we use beam search with a beam size of 10 and set the maximum generation length as 20.
\paragraph{Image-Text Retrieval.} We conduct experiments for both image-to-text retrieval (TR) and text-to-image retrieval (IR) on COCO ~\cite{Lin2014MicrosoftCC} and Flickr30K ~\cite{Plummer2015Flickr30kEC} datasets. We adopt the widely-used Karpathy split ~\citep{karpathy2015deep} for both COCO and Flickr30K. COCO contains 113k/5k/5k images for train/validation/test, and Flickr30K contains 29k/1k/1k images for train/validation/test. Following ~\cite{Li2021AlignBF, li2022blip}, we jointly optimize the ITC loss and the ITM loss during fine-tuning. During inference, we first select top-k candidates by computing the dot-product similarity between the image and text encoder features (When extracting the image encoder feature, for efficiency of coarse-grained ranking, we replace the TPD with a simple strategy in which we directly detect the patch in a transformer layer based on the self-attention weights of image [CLS] token to other patch tokens ), and then rerank the selected candidates based on their ITM scores (In the fine-grained reranking stage, for the same image, we re-extracting multiple image encoder features based on TPD with the guidance of multiple text candidates.). We set $k = 256$ for COCO and $k = 128$ for Flickr30K. 
\paragraph{Visual Grounding.} The task of visual grounding involves localizing the referred object in an image given a plain text query. Instead of directly regressing bounding boxes, our approach concatenates visual features with textual features, which are then fed into the multi-modal decoder to predict the object's coordinates. We evaluate our method on the referring expression grounding dataset: RefCOCO+\citep{yu2016modeling}. The RefCOCO+ dataset contains 19K images and 141K queries.

\end{document}